\documentclass[prd,preprint,floatfix,nofootinbib]{revtex4}
\usepackage{epsfig}
\usepackage{graphicx}
\usepackage{pstricks}

\def\lsim{\mathrel {\vcenter {\baselineskip 0pt \kern 0pt
    \hbox{$<$} \kern 0pt \hbox{$\sim$} }}}
\def\gsim{\mathrel {\vcenter {\baselineskip 0pt \kern 0pt
    \hbox{$>$} \kern 0pt \hbox{$\sim$} }}}
\def\slashchar#1{\setbox0=\hbox{$#1$}           % set a box for #1
 \dimen0=\wd0                                 % and get its size
  \setbox1=\hbox{/} \dimen1=\wd1               % get size of /
\ifdim\dimen0>\dimen1                        % #1 is bigger
  \rlap{\hbox to \dimen0{\hfil/\hfil}}      % so center / in box
  #1                                        % and print #1
  \else                                        % / is bigger
 \rlap{\hbox to \dimen1{\hfil$#1$\hfil}}   % so center #1
   /                                         % and print /
  \fi}                                         %
\def\cpto{\mathrel {\vcenter {\baselineskip 0pt \kern 0pt
    \hbox{$CP$} \kern 0pt \hbox{$\longrightarrow$} }}}
\def\cptof{\mathrel {\vcenter {\baselineskip 0pt \kern 0pt
    \hbox{$~CP$} \kern 0pt \hbox{$\longleftrightarrow$} }}}

\begin{document}

\baselineskip=15pt

\preprint{}

\title{Ansatz for small FCNC with a non-universal $Z^\prime$}

\author{Xiao-Gang He}
\email{hexg@phys.ntu.edu.tw}
\affiliation{Department of Physics and Center for Theoretical Sciences, \\
National Taiwan University, Taipei 106, Taiwan}

\author{G. Valencia}
\email{valencia@iastate.edu}
\affiliation{Department of Physics and Astronomy, Iowa State University, Ames, IA 50011, USA}

\date{\today $\vphantom{\bigg|_{\bigg|}^|}$}

\date{\today}

\vskip 1cm
\begin{abstract}

It is well known that a non-universal $Z^\prime$ induces tree-level FCNC which are severely constrained by experiment, most notably meson mixing. We point out that there is a class of models, with a down-quark mass matrix of the Georgi-Jarlskog form, in which the FCNC in the down-type quark sector vanish or are strongly suppressed. The largest FCNC in these models would occur in the $tc$ transition with a strength comparable to $V_{ts}$.

\end{abstract}

\pacs{PACS numbers: }

\maketitle

\section{Introduction}

Flavor changing neutral currents (FCNC) have been a powerful constraint on new physics for a long time now. After the recent round of flavor physics measurements at the B factories, there remains little room for flavor violation beyond that present in the CKM matrix. This has led to conjectures of ``minimal flavor violation'' MFV for theories beyond the SM.

In this paper we point out that it is possible to have new sources of flavor violation with little or no impact on existing measurements in flavor physics. This is illustrated with a specific ansatz for the down-type quark mass matrix in the context of non-universal $Z^\prime$ models.

A $Z^\prime$ boson appears in many extensions of the SM in different forms: an additional $U(1)$ symmetry; a linear combination of $U(1)$ and non-abelian gauge bosons (such as the $Z^\prime$ in left-right symmetric models);
a neutral gauge boson from a grand unified non-abelian symmetry model (such as $SO(10)$ and the string inspired $E_6$ models).
It is probably one of the best motivated extensions of the
SM. \cite{Langacker:2008yv}

Without specifying the underlying theory, the FCNC couplings in a non-universal $Z^\prime$ model are not known. However,  in models where the quark mass matrices are sufficiently constrained, the $Z^\prime$ FCNC couplings can be determined. Our ansatz requires that the theory has symmetric or hermitian mass matrices so that it is possible to relate the matrices that rotate the left and right handed quarks from the weak basis to the mass eigenstate basis. Examples of existing models which can naturally realize such mass matrices, are the manifest left-right symmetric model and the SO(10) grand unification model.  It is also necessary to have sufficient information  about the up and down quark mass matrices so that it is possible to reconstruct the unitary matrices which diagonalize them from the experimentally known CKM matrix elements. A down-type quark mass matrix of the Georgi-Jarlskog form is one such example.  In the examples we consider, the largest FCNC coupling occurs in the $t\to c$ transition and is comparable to $V_{ts}$.

\section{General $Z^\prime$ couplings to fermions}

We begin by writing down a general interaction Lagrangian between a $Z^\prime$ and quarks.
We will initially assume that this interaction is diagonal in the weak basis but not necessarily universal:
\begin{eqnarray}
{\cal L}_{Z^\prime} = {g\over 2 \cos\theta_W} (\bar U_L \delta^U_L \gamma_\mu U_L +
\bar U_R \delta^U_R \gamma_\mu U_R + \bar D_L \delta^D_L \gamma_\mu D_L +
\bar D_R \delta^D_R \gamma_\mu D_R) Z^{\prime\mu}\,
\end{eqnarray}
where $U_{L,R} = (u, c, t)^T_{L,R}$, $D = (d, s, b)^T_{L,R}$. $\delta^{U,D}_{L,R} = diag(\Delta^{u,d}_{L,R}, \Delta^{c,s}_{L,R}, \Delta^{t,b}_{L,R})$.

In the mass eigenstate basis, $q^{U,D}_{L,R} =V^{U,D}_{L,R}q^{m, U,D}_{L,R}$, the interaction looks like
\begin{eqnarray}
{\cal L}_{Z^\prime} &=& {g\over 2 \cos\theta_W} (\bar U_L V^{U\dagger}_L\delta^U_LV^U_L \gamma_\mu U_L +
\bar U_R V^{U\dagger}_R\delta^U_R V^U_R\gamma_\mu U_R \nonumber\\
&+& \bar D_L V^{D\dagger}_L\delta^D_L V^D_L \gamma_\mu D_L +
\bar D_R V^{D\dagger}_L \delta^D_R V^D_R \gamma_\mu D_R) Z^{\prime^\mu}\,
\end{eqnarray}
It is then clear that FCNC will be induced unless the $\delta^i_j$ are proportional to the unit matrix, that is, unless the $Z^\prime$ couplings are universal.

In general, in non-universal $Z^\prime$ models, there are tree-level FCNC  \cite{Langacker:2000ju}. Some of these models that have been studied recently single out the couplings of the third generation \cite{He:2002ha,He:2003qv,Barger:2009eq,Barger:2009qs}. This feature can be represented in general by setting $\Delta^i_j=0$ above, except for $\Delta^{t,b}_{L,R}=\kappa^{t,b}_{L,R}$ with $\kappa^{t,b}_{L,R}$ parametrizing the strength of the new interaction,
\begin{eqnarray}
{\cal L}_{Z^\prime} = - {g\over 2\cos\theta_W} \,
\left( \kappa_L^t \bar{t}\gamma^\mu P_L t  + \kappa_L^b \bar{b}\gamma^\mu P_L b + \kappa_R^t \bar{t}\gamma^\mu P_R t + \kappa_R^b \bar{b}\gamma^\mu P_R b\right)
Z^{\prime_\mu},
\label{zpcoup}
\end{eqnarray}
and possibly with corresponding couplings to the first two generations that are suppressed by a small parameter $r$ with respect to Eq.~\ref{zpcoup}. Phenomenologically, the couplings to the first two generations must be essentially the same.

FCNC in the down-quark and up-quark sectors still occur in the mass eigenstate basis, and are given by
\begin{eqnarray}
&&{\cal L}_{FCNC} = {g\over 2\cos\theta_W}\, \left( \bar{U}_i \gamma^\mu \left(\kappa^t_L a^u_{ij}P_L+\kappa^t_Rb^u_{ij}P_R\right)U_j  \, +\, \bar{D}_i \gamma^\mu \left(\kappa^b_L a^d_{ij}P_L+\kappa^b_Rb^d_{ij}P_R\right)D_j \right) Z^{\prime_\mu}\nonumber \\
&&a^{d}_{ij} =  V^{D\dagger}_L Z_{\kappa} V^D_L \;,\;\;a^{u}_{ij} = V^{U\dagger}_L Z_{\kappa} V^U_L\;,\;\;b^{d}_{ij} =  V^{D\dagger}_R Z_{\kappa} V^D_R \;,\;\;b^{u}_{ij} = V^{U\dagger}_R Z_{\kappa} V^U_R,
\label{fcncint}
\end{eqnarray}
in terms of the matrix $Z_{\kappa} = {\rm diag}(r,r,1)$.

In general, models with an extra $U(1)$ may have FCNC couplings  already in the weak basis, but these do not exist for suitable  quantum number assignments.  Models with $Z^\prime$ couplings  to fermions proportional to $Z_{\kappa}$ in the weak interaction basis can be constructed easily. For example, if one
assumes that in the manifest left-right symmetric model, the third generation transforms under an additional $U(1)_{3rd}$ symmetry. If the gauge boson of this gauge symmetry is the $Z^\prime$, its couplings to fermions have the desired  form proportional to $Z_{\kappa}$.

\section{Constraints on FCNC}

Flavor physics processes are known to severely constrain FCNC of the type in Eq.~\ref{fcncint} and these have been analyzed recently in the context of non-universal $Z^\prime$ models. Generally it was found that the most severe constraints at present arise from $K$, $D$, $B_d$ and $B_s$ meson mixing.\footnote{Many other processes have also been used to constrain the model parameters \cite{He:2004it,He:2006bk,He:2007iu,Barger:2009eq,Barger:2009qs,Chiang:2006we,Baek:2006bv,Baek:2008vr,Chen:2008za, Mohanta:2008ce}, and stringent bounds have been obtained for the FCNC couplings.}
For example, when the new couplings are purely right handed, the interactions in Eq.~\ref{fcncint} induce new physics contributions of the form
\begin{eqnarray}
\Delta M_{ij} \sim \frac{M_Z^2}{M^2_{Z^\prime}}( \kappa^{t,b}_R)^2 (b^{u,d}_{ij})^2.
\label{mix}
\end{eqnarray}
The resulting constraints can be summarized as follows \cite{He:2004it,He:2006bk,He:2007iu}
 \begin{eqnarray}
 \left(\frac{M_Z}{M_{Z^\prime}}\, \kappa_b\right) b_{ij}^d \lsim \left( \matrix {- &  10^{-4} &  10^{-4} \nonumber \\
                            10^{-4} & - & 10^{-3} \nonumber\\
                            10^{-4} & 10^{-3}  & - }
                            \!\!\!\!\!\!\!\!\!\!\!\!\!\!\!\!\!\!\! \right),\,\,
\left( \frac{M_Z}{M_{Z^\prime}}\, \kappa_t\right) b_{ij}^u \lsim \left( \matrix {- &  10^{-4} &  ? \nonumber \\
                            10^{-4} & - &  ?\nonumber\\
                            ? & ?  & - }
                            \!\!\!\!\!\!\!\!\!\!\!\!\!\!\!\!\!\!\! \right).
                            \label{numbounds}
\end{eqnarray}
where the question marks reflect the lack of knowledge about FCNC for the top-quark.

For the specific models discussed in Ref.~\cite{He:2002ha,He:2003qv}, the overall strength of the interaction, $(M_Z/M_{Z^\prime})\kappa_R^{b,t}$, is approximately one. Clearly, if the interaction is weaker, the resulting constraints on $b^{d,u}_{i,j}$ are also weaker. The upper bound for $b^u_{12,21}$ in Eq.~\ref{numbounds} arises from attributing all the observed $D-\bar{D}$ mixing to FCNC \cite{He:2007iu}, and this is obviously an overestimate.

Numerical constraints for the more general case in which the $Z^\prime$ has both left and right handed interactions can be found in the literature. The meson mixing induced in the more general case has a more complicated form than Eq.~\ref{mix}, but numerically, the resulting constraints on $a^{d,u}_{i,j}$ are the same order of magnitude as those on $b^{d,u}_{i,j}$ and both are well represented  by Eq.~\ref{numbounds}.

\subsection{Models where $V^{d,u}_{L,R}$ are known}

The mixing matrices $V^{d,u}_{L,R}$ are related to the CKM matrix $V_{CKM} = V^{u\dagger}_LV^d_L$. Since the experimentally measurable quantities are directly related to the elements in $V_{CKM}$, it is not possible in general to separately extract the elements  in $V_{L}^{u,d}$.  Furthermore, since $V_{R}^{u,d}$ does not play a role in general in $V_{CKM}$, the elements in $V_R^{u,d}$ are not known either. Our purpose in this paper is to point out that in certain models the left and right-handed rotation matrices $V^{d,u}_{L,R}$ are related and there is sufficient information to predict the
$Z^\prime$ FCNC couplings.

In particular, if the fermion mass matrices $M^i$ are Hermitian, they can be diagonalized by the transformation
\begin{eqnarray}
 M^i = V_L^{i}  \hat M^i V_R^{i\dagger} = V^i_L \hat M^i V^{i\dagger}_L\;,
\end{eqnarray}
where $\hat M^i$ indicates a diagonal mass matrix. Therefore, in this class of models $V^i_R$ is determined to be equal to $V^i_L$ up to possible phases.
This class of models occurs naturally  in left-right  models with manifest left-right symmetry.

Relations between $V_L$ and $V_R$ also exist in the case of  symmetric mass matrices, $M^{iT} = M^i$, as occurs in SO(10) grand unification models. In this case
\begin{eqnarray}
M^i = V_L^{i} \hat M^i V_R^{i\dagger} = V_L^{i} \hat M^i V_L^{iT}\;,
\end{eqnarray}
and therefore $V_{R} = V_L^*$.

In order to know separately $V^{u}_{L,R}$ and $V^d_{L,R}$ and therefore fix the FCNC $Z^\prime$ couplings, additional information about $M^{u,d}$ is needed. A simple possibility is that one of the rotation matrices $V_{L}^{u}$ or $V_L^d$ is the unit matrix. That is, that either the up or down type quark mass matrix is already diagonal in the weak eigenstate basis. In this case the other rotation matrix is completely determined in terms of the CKM matrix. For the case of a hermitian mass matrix we have the two possibilities,
\begin{eqnarray}
&&a)V_L^D=I: \;\; a^u_{ij}(b_{ij}^u) =  V_{CKM} Z_{\kappa} V^\dagger_{CKM}\;,\;\;a^d_{ij}(b_{ij}^d)
=  Z_{\kappa}\;,\nonumber\\
&&b)V_L^U=I: \;\; a^d_{ij}(b_{ij}^d) = V^{\dagger}_{CKM} Z_{\kappa}
V_{CKM}\;,\;\;a^u_{ij}(b_{ij}^u) =  Z_{\kappa}
 \;.
\end{eqnarray}
Replacing $b_{ij}^{d,u} = a^{d,u*}_{ij}$, one obtains the couplings for symmetric mass matrix case.

Explicitly, we have
\begin{eqnarray}
V_{CKM}^\dagger Z_{\kappa} V_{CKM}  = \left( \matrix {r & {\cal O}(\lambda^5) & A(1-r) \lambda^3(1-\rho+i\eta) \nonumber \\
                           {\cal O}(\lambda^5) & r & -A(1-r)\lambda^2 \nonumber\\
                            A (1-r)\lambda^3 (1-\rho - i\eta) & -A(1-r)\lambda^2  & 1 }
                           \!\!\!\!\!\!\!\!\!\!\!\!\!\!\!\!\!\!\! \right),
\end{eqnarray}
\begin{eqnarray}
V_{CKM} Z_{\kappa} V_{CKM}^\dagger  = \left( \matrix {r & {\cal O}(\lambda^5) & A(1-r) \lambda^3(\rho- i\eta) \nonumber \\
                           {\cal O}(\lambda^5) & r & A(1-r)\lambda^2 \nonumber\\
                            A (1-r)\lambda^3 (\rho + i\eta) & A(1-r)\lambda^2  & 1 }
                           \!\!\!\!\!\!\!\!\!\!\!\!\!\!\!\!\!\!\! \right)\;.
\end{eqnarray}
Numerically, this makes scenario $(a)$ compatible with the constraints of Eq.~\ref{numbounds}, and scenario $(b)$ incompatible in general. Scenario $(b)$ can be reconciled with experimental constraints in two cases: for models where the non-universality of the $Z^\prime$ is ``small'' ($r$ is close to one); or for models where the overall coupling strength $(M_Z /M_{Z'})\kappa_b$ is substantially smaller than 1.

It is also possible to find examples with non-trivial mass matrices satisfying the FCNC constraints. There are many attempts in the literature to obtain predictive quark mass matrices \cite{Georgi:1979df,Cheng:1987rs,He:1989eh,Fritzsch:1999ee} and therefore known forms for $V_{L,R}$. Among them, we note that the Georgi-Jarlskog mass matrix for the down-quark sector \cite{Georgi:1979df}  has the form needed to produce a FCNC $Z^\prime$ matrix  consistent with the phenomenological constraints discussed above.  A mass matrix $M_D$ of the Georgi-Jarlskog form\footnote{which reproduces the relation $\sqrt{m_d/m_s}\sim \lambda$ with $B=A\lambda/(1-\lambda^2)$}  is diagonalized by the matrix $V^d_L$, where
\begin{eqnarray}
M_D \sim \left( \matrix {0 & B & 0 \nonumber \\
                            B & A & 0 \nonumber\\
                            0 & 0  & C }
                           \!\!\!\!\!\!\!\!\!\!\!\!\!\!\!\!\!\!\! \right)\,, \,\,
V^d_L  \sim \left( \matrix {1 & \lambda & 0 \nonumber \\
                            -\lambda & 1 & 0 \nonumber\\
                            0 & 0  & 1 }
                           \!\!\!\!\!\!\!\!\!\!\!\!\!\!\!\!\!\!\! \right)
 \, .
\end{eqnarray}
Since $V^d_L=V^d_R$ in this type of models, one easily sees that
$b_{ij}^d = Z_\kappa$,
naturally satisfying all FCNC constraints in the down-type sector. It is interesting to point out that if the couplings of the $Z^\prime$ to the first two generations are different, say parametrized by $r_1$ and $r_2$ in the $11$ and $22$ entries in $Z_\kappa$, then
\begin{eqnarray}
b^d_{sd}\sim (r_1-r_2)\lambda
\end{eqnarray}
and satisfying the constraints, Eq.~\ref{numbounds}, requires the first two generations to have essentially the same couplings to the $Z^\prime$, $(r_1-r_2) \lsim 5\times 10^{-4}$.

We can then turn our attention to the up-quark sector where, unfortunately, the Georgi-Jarlskog ansatz does not reproduce successfully  the CKM matrix and must be modified. From a purely phenomenological perspective we are interested in a rotation matrix $V^u_L=V^u_R (V_R^*)$ such that $V^{u\dagger}_L V^d_L = V_{CKM}$. This is accomplished by a matrix of the form
\begin{eqnarray}
V^u_L  \sim \left( \matrix {1 & -b \lambda^2 & -A \lambda^3(\rho-i\eta) \nonumber \\
                            b \lambda^2 & 1 & -A\lambda^2 \nonumber\\
                            A \lambda^3 (\rho + i\eta) & A\lambda^2  & 1 }
                           \!\!\!\!\!\!\!\!\!\!\!\!\!\!\!\!\!\!\! \right)
 \,\, .
\end{eqnarray}
With $V^u_L=V^u_R $ we are thus led to
\begin{eqnarray}
b_{ij}^u \sim \left( \matrix {r  & {\cal O}(\lambda^5) & (1-r)A\lambda^3(\rho-i\eta) \nonumber \\
                       {\cal O}(\lambda^5)       &r  & (1-r)A\lambda^2 \nonumber\\
                            (1-r)A\lambda^3(\rho+i\eta)  &  (1-r)A\lambda^2 & 1 }
                           \!\!\!\!\!\!\!\!\!\!\!\!\!\!\!\!\!\!\! \right)
 \,\, ,
\end{eqnarray}
this case has the same FCNC mediated by $Z^\prime$ in the up quark
sector as scenario $(a)$ above. In both cases, the $c\to u$ flavor
changing transition is at the level corresponding to the upper bound
derived from $D-\bar{D}$ mixing.

Both scenarios satisfying the constraints in the down-quark sector predict that the largest FCNC, and perhaps the only one large enough to be observed, occurs in the $t\to c$  transition at the level of $V_{ts}$. A quick estimate suggests that this will be difficult to observe at LHC. The analysis of
Ref.~\cite{Han:1995pk} parametrizes the anomalous FC $Z$ coupling in the form
\begin{eqnarray}
{\cal L} = -{g\over 2 \cos\theta_W} \kappa^R_{tc} (\bar t  \gamma_\mu P_R c) Z^\mu + {\rm ~h.c.},
\label{peccei}
\end{eqnarray}
and finds that with 100~fb$^{-1}$, the LHC can reach a $99\%$~c.l. sensitivity  $\kappa^R_{tc} \sim 0.02$.  This result arises from the study of the branching ratio for the mode $t\to Zc$, and corresponds to a lowest observable branching ratio $B(t\to Z c) \sim {\rm ~few~}\times 10^{-4}$. A coupling as in Eq.~\ref{peccei} is induced in non-universal $Z^\prime$ models via the $Z-Z^\prime$ mixing parameter $\xi_Z$
\begin{eqnarray}
\kappa^R_{tc} \sim \kappa A \lambda^2 \xi_Z \sim {\rm ~few~} \times 10^{-3}.
\end{eqnarray}
Where the last number follows because the parameter $\xi_Z$ is typically of order $(M_Z/M_{Z^\prime})^2$ \cite{Langacker:2000ju,He:2002ha}.

These numbers show that it is unlikely that this FCNC can be observed at LHC in top-quark decay. However, the top quark decay process is suppressed by the $Z-Z^\prime$ mixing parameter in addition to the intrinsic FCNC strength because the $Z^\prime$ is too heavy to be  produced directly. One could  search instead for $B(t\to c \ell^+\ell^-)$ via a $Z^\prime$ exchange, but the results are also too small to be observed, of ${\cal O}(10^{-6})$ for the most optimistic scenario where $\ell=\tau$ and the couplings to third generation leptons are also enhanced. Another possibility to observe this coupling would be in the single-top production process, in a manner similar to that discussed in Ref.~\cite{Arhrib:2006sg} for LHC and ILC.

\section{Conclusions}

We have presented simple ans\"atze for the down-type quark mass matrix that naturally suppress the FCNC introduced in non-universal $Z^\prime$ models to levels compatible with existing constraints. These models then predict that the largest FCNC transitions occur in the $t\to c$ with a strength comparable to $V_{ts}$.

Finally we note that even though the flavor physics measurements are in substantial agreement with the standard model, there are a few observables deviating by $2-3\sigma$ from the standard model expectation, the phase in $B_s$ mixing being one example. A more ambitious program would consist of modifying our ans\"atze in an attempt to explain such deviations. This would require specifying the mass matrices with much greater detail than we have done here, but would be premature at this stage.

\begin{acknowledgments}

The work of X.G.H.  was supported in part by NSC and NCTS.
The work of G. V. was supported in part by DOE under contract number DE-FG02-01ER41155. G. V. thanks the Department of Physics at National Taiwan University for its hospitality while this work was completed.

\end{acknowledgments}

\appendix

\end{document}